\documentclass[aps,prl,twocolumn,superscriptaddress,showpacs]{revtex4-1}
\usepackage{graphicx}
\usepackage{dcolumn}
\usepackage{bm}
\usepackage{color}

\begin{document}
\preprint{APS/123-QED}

\title{Evidence for weakly correlated oxygen holes in the highest-T$_{c}$ cuprate superconductor HgBa$_2$Ca$_2$Cu$_3$O$_{8+\delta}$}

\author{A. Chainani}
\affiliation{RIKEN SPring-8 Centre, 1-1-1 Sayo-cho, Hyogo 679-5148, Japan}
\affiliation{Institut Jean Lamour, Universit\'{e} de Lorraine, UMR 7198 CNRS, BP70239, 54506 Vandoeuvre l\'{e}s Nancy, France}
\author{M. Sicot}
\affiliation{Institut Jean Lamour, Universit\'{e} de Lorraine, UMR 7198 CNRS, BP70239, 54506 Vandoeuvre l\'{e}s Nancy, France}
\author{Y. Fagot-Revurat}
\affiliation {Institut Jean Lamour, Universit\'{e} de Lorraine, UMR 7198 CNRS, BP70239, 54506 Vandoeuvre l\'{e}s Nancy, France}
\author{G. Vasseur}
\affiliation{Institut Jean Lamour, Universit\'{e} de Lorraine, UMR 7198 CNRS, BP70239, 54506 Vandoeuvre l\'{e}s Nancy, France}
\author{J. Granet}
\affiliation{Institut Jean Lamour, Universit\'{e} de Lorraine, UMR 7198 CNRS, BP70239, 54506 Vandoeuvre l\'{e}s Nancy, France}
\author{B. Kierren}
\affiliation{Institut Jean Lamour, Universit\'{e} de Lorraine, UMR 7198 CNRS, BP70239, 54506 Vandoeuvre l\'{e}s Nancy, France}
\author{L. Moreau}
\affiliation{Institut Jean Lamour, Universit\'{e} de Lorraine, UMR 7198 CNRS, BP70239, 54506 Vandoeuvre l\'{e}s Nancy, France}
\author{M. Oura}
\affiliation{RIKEN SPring-8 Centre, 1-1-1 Sayo-cho, Hyogo 679-5148, Japan}
\author{A. Yamamoto}
\affiliation{Strong Correlation Physics Division, RIKEN Center for Emergent Matter Science (CEMS), Wako 351-0198, Japan} 
\author{Y. Tokura}
\affiliation{Strong Correlation Physics Division, RIKEN Center for Emergent Matter Science (CEMS), Wako 351-0198, Japan}
\author{D. Malterre}
\affiliation{Institut Jean Lamour, Universit\'{e} de Lorraine, UMR 7198 CNRS, BP70239, 54506 Vandoeuvre l\'{e}s Nancy, France}

\date{\today}

\begin{abstract}

We study the electronic structure of  HgBa$_2$Ca$_2$Cu$_3$O$_{8+\delta}$ (Hg1223 ; T$_{c}$ = 134 K) using photoemission  spectroscopy (PES) and x-ray absorption spectroscopy (XAS). Resonant valence band PES across the O K-edge and Cu L-edge identify correlation satellites originating in O 2p and Cu 3d two-hole final states, respectively. Analyses using the experimental O 2p and Cu 3d partial density of states show quantitatively different on-site Coulomb energy for the Cu-site (U$_{dd} =~$6.5$\pm$0.5 eV) and O-site (U$_{pp}$ =~1.0$\pm$0.5 eV). Cu$_{2}$O$_{7}$-cluster calculations with non-local screening explain the Cu 2p core level PES and Cu L-edge XAS spectra, confirm the U$_{dd}$ and U$_{pp}$ values, and provide evidence for the Zhang-Rice singlet state in Hg1223. In contrast to other hole-doped cuprates and 3d-transition metal oxides, the present results indicate weakly correlated oxygen holes in Hg1223. 

\end{abstract}


\maketitle

Nearly 30 years after the discovery of high-transition temperature (T$_{c}$) superconductivity in the cuprates,\cite{Bed}
the role of electron correlations still remains the central enigma in understanding their properties.
While pioneering theoretical and experimental studies have established important aspects of spin-charge ordering,\cite{ZG,Machida,JT,Salkola,Wu,GG,JC,MT,Tabis,HJ} anti-ferromagnetic correlations\cite{Julien,Lee,Chan} and electron-phonon coupling\cite{Lanzara} in the high-T$_c$ cuprates, the origin for high-T$_{c}$ superconductivity still remains elusive.\cite{Keimer} The spin-charge order competes with the metallicity of doped carriers in the CuO${_2}$ layers and leads to novel thermodynamic, transport and spectroscopic phenomena which favor quantum critical behavior.\cite{Valla,Randeria,Pepin,Jacobs,Proust} 

It is well-accepted that Mott-Hubbard correlations provide an appropriate starting point for describing the electronic structure of the CuO$_{2}$ layers.\cite{Emery,Varma,Zhang,arpesRMP,MottRMP,Weber,Werner} However, the actual quantification of correlations in terms of an on-site Coulomb energy for the Cu-site(=U$_{dd}$) in comparison with that on the O site(=U$_{pp}$), and their roles in spin-charge order and superconductivity is still being investigated. In fact, using the method of Cini\cite{Cini} and Sawatzky\cite{Sawatzky} (see supplementary material SM for a description of the method), it is known that U$_{pp}$ is large ($\sim$5-6 eV) and quite comparable to U$_{dd}$ ($\sim$ 5-8 eV) in YBa$_{2}$Cu$_{3}$O$_{7}$ (YBCO)\cite{Marel,Balzarotti}, Bi$_{2}$Sr$_{2}$CaCu$_{2}$O$_{8}$(Bi-2212),\cite{Tjeng} and La$_{2-x}$A$_{x}$CuO$_{4}$(A = Sr, Ba).\cite{BarDeroma} This behaviour of U$_{pp}$ $\sim$ U$_{dd}$ is known for oxides across the 3d transition metal(TM) series : titanium/vanadium oxides(SrTiO$_{3}$, V$_{2}$O$_{3}$, VO$_{2}$, V$_{2}$O$_{5}$),\cite{Ishida,Post,Park} LaMO$_{3}$ (M = Mn-Ni) perovskites,\cite{AC,DD} and cuprates(including Cu$_{2}$O and CuO).\cite{Marel,Balzarotti,Ghijsen,Tjeng,BarDeroma} A recent theoretical study on rare-earth nickelates (RNiO$_{3}$), using comparable values of U$_{dd}$ (= 7 eV) and U$_{pp}$ (= 5 eV), showed that the metal-insulator transition in RNiO$_{3}$ arises from a novel charge-order involving ligand holes.\cite{Johnston} It would be interesting to establish the relation of U$_{dd}$ and U$_{pp}$ with charge-order even for the cuprates. However, while all the above mentioned cuprate families
show charge-order,\cite{GG,JC,MT,Tabis,HJ} to date, there is no report of charge-order in HgBa$_2$Ca$_2$Cu$_3$O$_{8+\delta}$(Hg1223) series, which shows the highest-T$_{c}$(= 134 K) at ambient pressure,\cite{Ott} although a clear spin gap below T* $\sim$230 K was reported early\cite{Julien}. Given the recent resurgence of interest in the Hg-based cuprates,\cite{Yamamoto,Loret,Manuel,Jang,Hinton} we felt it important to quantify U$_{dd}$ and U$_{pp}$ for the optimally doped Hg1223. This would help to understand the highest T$_c$ system, motivate studies for charge order in Hg1223 and clarify its link with U$_{dd}$ and U$_{pp}$ in general.     
 
Photoemission spectroscopy has played a vital role in elucidating the electronic structure of high-T$_c$ cuprates. High-resolution angle-resolved PES studies have revealed the Fermi surfaces,\cite{arpesRMP} a momentum-dependent pseudogap,\cite{Marshall,Ding} a low energy kink in the band dispersion,\cite{Lanzara} and the highly intriguing Fermi arcs\cite{Norman} whose origin is still under debate.\cite{Meng,King} Core level PES in combination with model cluster calculations showed that U$_{dd}$ is larger than the charge-transfer energy $\Delta$ and the low energy properties involve charge-transfer excitations in the Zaanen-Sawatzky-Allen scheme.\cite{Fujimori,ZXS} The properties of all the cuprate families originate in the CuO$_{4}$ plaquette common to their layered structure.\cite{arpesRMP} The strong hybridization of the Cu 3d-O 2p states leads to the Zhang-Rice singlet(ZRS) as the lowest energy state.\cite{Zhang} For Bi2212 \cite{Veenendaal94,Brookes2001} and for LSCO\cite{Taguchi,Brookes}, core level PES and spin-polarized resonant PES studies have established the ZRS state across the superconducting dome. Since the highest T$_{c}$ Hg1223 system also shows the superconducting dome behavior,\cite{Fukuoka,Yamamoto} it is important to identify and characterize the ZRS state in Hg1223.

We carry out XAS and resonant PES across the O K-edge and Cu L-edge of optimally doped Hg1223. The samples were prepared using a high-pressure method and characterized for their superconducting T$_{c}$ = 134 K (Fig. S1), as reported recently.\cite{Yamamoto} The details of sample preparation, characterization and spectroscopy measurements are described in SM.\cite{SM} We also carry out off-resonant VB-PES at specific photon energies from 21.2 eV-1200 eV to conclusively determine the O 2p and Cu 3d pDOS. Using the estimated U$_{dd}$ and U$_{pp}$, we carry out model many-body Hamiltonian calculations for a Cu$_{2}$O$_{7}$-cluster with non-local screening\cite{Veenendaal,Okada} for the Cu 2p core level PES and Cu L-edge XAS spectra. The results show the importance of non-local screening in PES and XAS, confirm the ZRS state as well as the estimated values of  U$_{dd}$ = 6.5$\pm$1 eV and U$_{pp}$ = 1.0$\pm$0.5 eV. We conclude that the highest T$_{c}$ cuprate exhibits an unexpectedly smaller U$_{pp}$ compared to U$_{dd}$ i.e. coexistence of weakly correlated oxygen holes and strongly correlated d-holes, which need to be considered in theoretical models addressing the highest T$_{c}$ in cuprates.  

\begin{figure}
\centering
\includegraphics[width=\columnwidth]{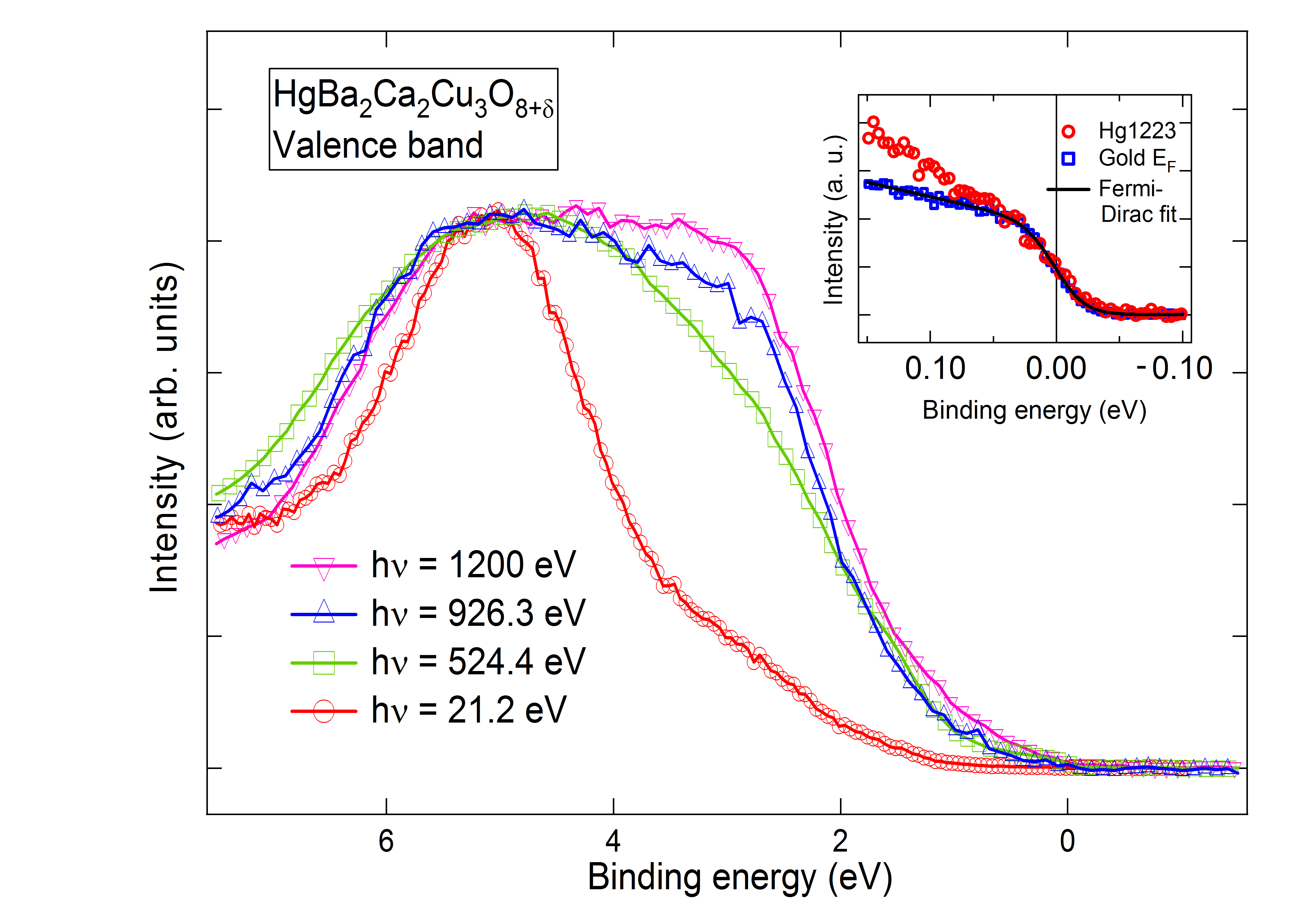}
\caption{\label{las} {(color online) Photon energy dependent valence band PES of HgBa$_2$Ca$_2$Cu$_3$O$_{8+\delta}$(Hg1223) measured using h$\nu$ = 21.2 eV, 524.4 eV, 926.3 eV and 1200 eV.
The spectra are normalized at 5 eV BE. Inset shows the Fermi step for Hg1223 and gold reference
along with a fit to the Fermi-Dirac distribution convoluted with a Gaussian function. The He I (h$\nu$ = 21.2 eV) spectrum is a laboratory measurement at T = 120 K and all other spectra are measured at T = 20 K using synchrotron radiation.}}
\end{figure}

Fig. 1 shows the valence band photoemission spectra of Hg1223 measured with ultraviolet (21.2 eV) and soft x-ray ($\sim$524, $\sim$926 eV and 1200 eV) photons. The valence band PES of Hg-1223 measured with a photon energy of 21.2 eV (He I) shows 
a broad peak at 5 eV binding energy(BE) and a weak feature at 2.5 eV extending to the Fermi level, consistent with earlier work.\cite{Almeras,Vasquez} Inspite of the very low intensity near E$_{F}$, the Hg1223 spectra show a Fermi step indicative of a metal. The near E$_F$ Hg1223 spectrum with He I photons measured at T = 120 K is shown in the inset along with the Fermi step for gold and a Fermi-Dirac fit to the step. The superconducting gap at T = 120 K is too small to be measured with a resolution of $\sim$40 meV. The feature at 5 eV BE is the O 2p band and the weak feature at 2.5 eV dominantly consists of the Cu 3d states, based on band structure calculations.\cite{Rodriguez} On increasing photon energies, the feature at 2.5 eV is relatively enhanced (the spectra are normalized at the 5 eV feature), confirming the assignment. The core-levels ( Hg 4f, Ba 3d, Ca 2p and O 1s) and wide BE range valence band spectrum showing also the Hg 5d, Ba 5s-5p, Ca 3p and O 2s shallow core levels measured with h$\nu$ = 1200 eV are discussed in SM(Figs. S2 and S3).

\begin{figure}
\centering
\includegraphics[width=\columnwidth]{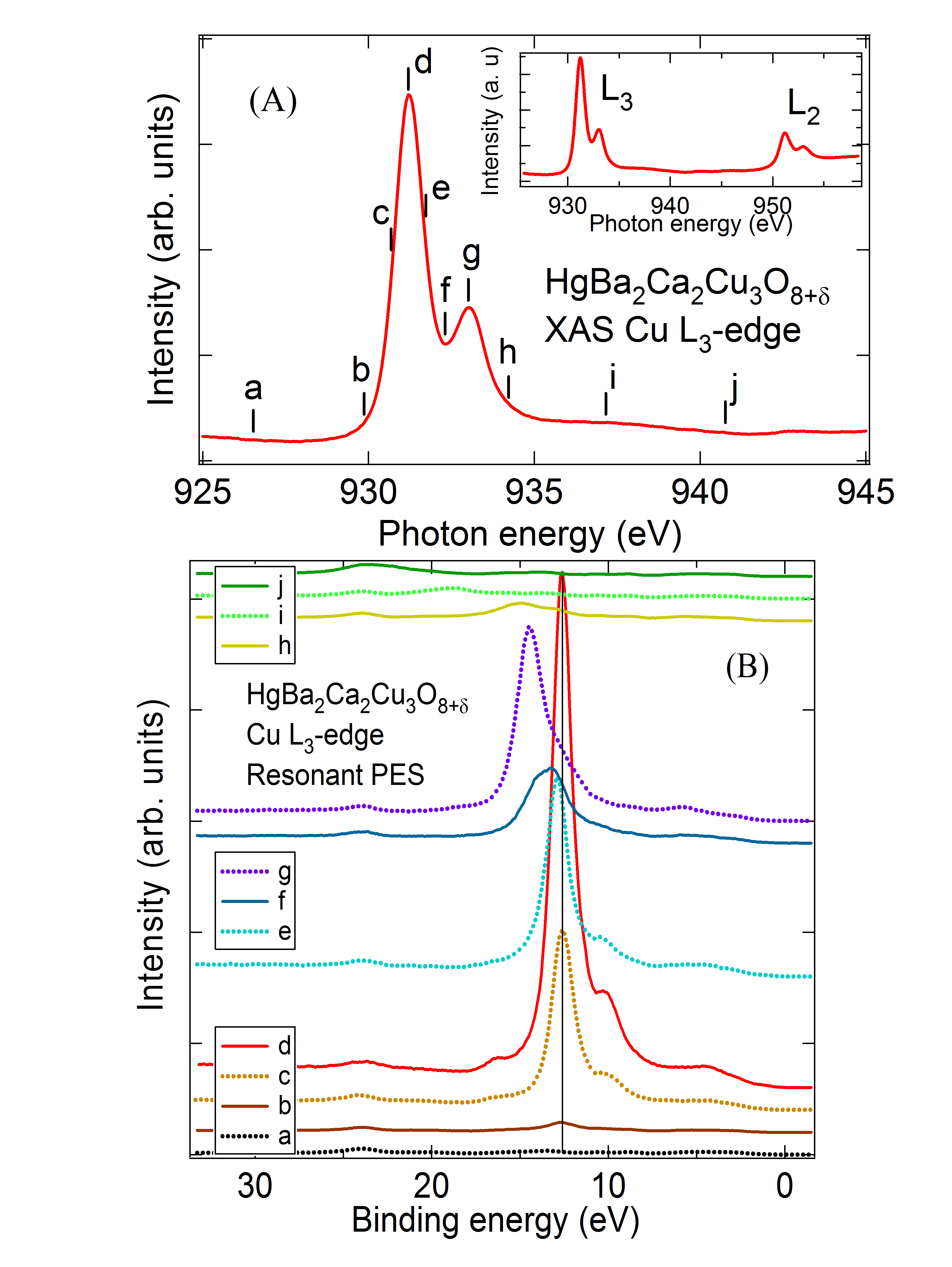}
\caption{\label{las} {(color online)(A) The Cu L$_{3}$-edge(2p-3d) XAS of Hg1223 measured at T = 20 K. The labels a-j indicate h$\nu$ used for measuring Cu 2p-3d resonant valence band PES of Hg1223 shown in panel (B). 
Resonant PES spectra are normalized for the incident photon flux and number of scans. The sharp peak at 12.6 eV BE obtained with h$\nu$ = 931.2 eV (main peak in XAS ; label $\bf{d}$ in panel A) is the 3d$^8$ giant resonance feature. The thin black vertical line shows that the 3d$^8$ resonance peak stays fixed at 12.8 eV BE upto h$\nu$ = 931.2 eV and evolves into the Cu L$_{3}$VV Auger satellite for higher h$\nu$.}}
\end{figure}

In Fig.2(A), we show the Cu L$_{3}$-edge XAS (inset shows the wide range L$_{3}$ and L$_{2}$ features), which consists of a main peak at $\sim$931 eV photon energy (label $\bf{d}$) and a weak but clear feature at $\sim$933 eV (label $\bf{g}$). The weak feature has been reported earlier but with much lower relative intensity.\cite{Pellegrin} We then carried out Cu 2p-3d resonant valence band PES at the photon energies labelled $\bf{a-j}$ and the spectra are correspondingly labelled and shown in Fig. 2(B). The off-resonance spectrum $\bf{a}$ is the same as the h$\nu$ = 926.3 eV spectrum of Fig. 1. As we increase the photon energies from $\bf{a}$ to $\bf{d}$, we see a dramatic $\sim$190-times increase in the intensity of  feature centered at 12.8 eV BE. This corresponds to the well-known giant resonance seen in cuprates.\cite{Tjeng} On increasing the photon energy further, we see a shift of the resonance feature as it moves to higher binding energies, with the magnitude of the energy shift tracking the increase in photon energy. This identifies the resonance feature as the Cu L$_{3}$VV Auger state, with two final state holes(VV) in the valence band thus confirming its origin to be the Cu 3d correlation satellite. The intensity follows the XAS profile with a reduction for photon energies labelled $\bf{e}$ and $\bf{f}$, then an increase at $\bf{g}$, followed by a gradual reduction at higher energies. 

\begin{figure}
\centering
\includegraphics[width=\columnwidth]{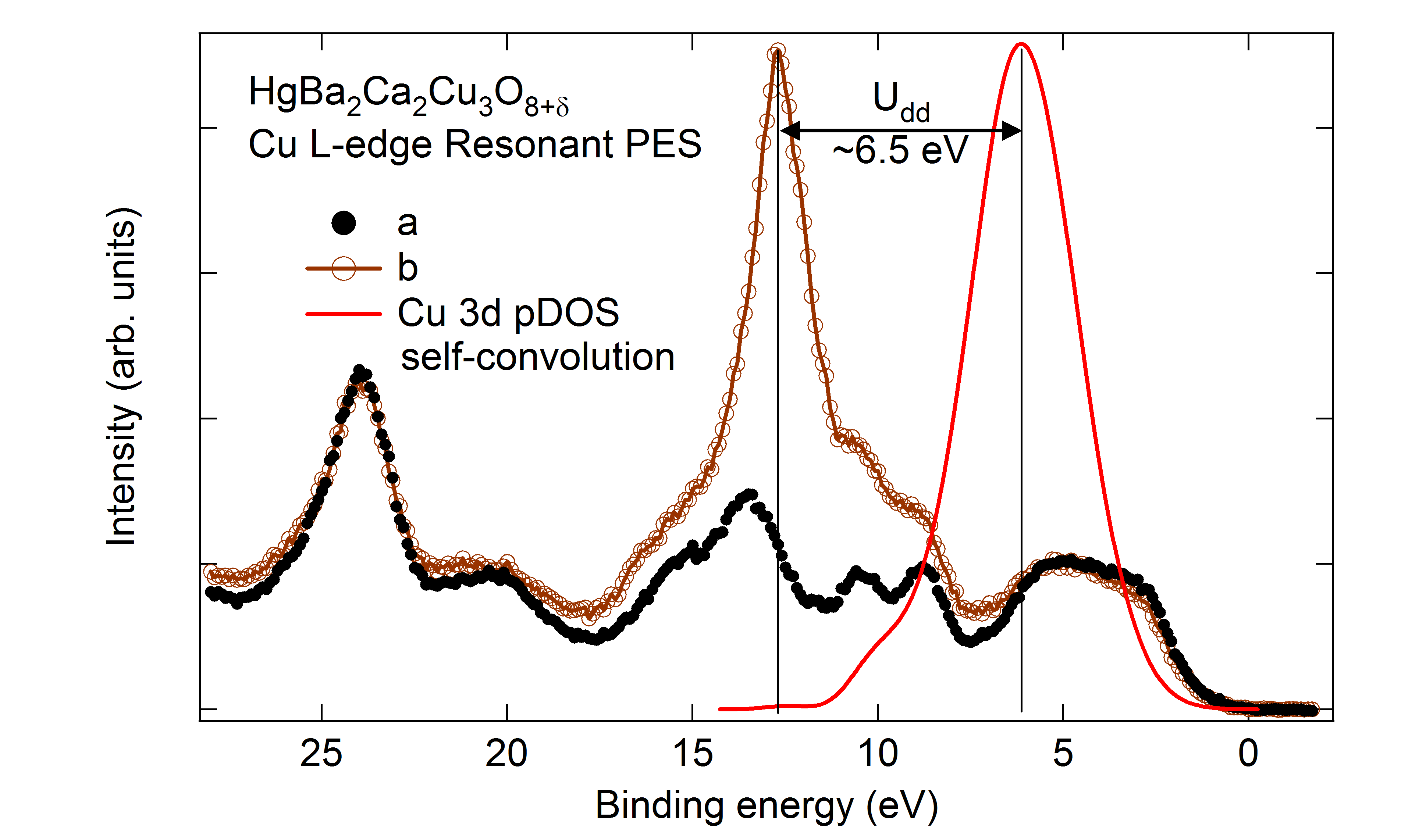}
\caption{\label{las} {(color online) (A) The Cu 2p-3d resonant PES spectra (labelled $\bf{a}$ and $\bf{b}$, from Fig. 2(B)) of Hg1223 measured at T = 20 K (normalized at the O 2s feature at $\sim$ 24 eV BE) showing the resonantly enhanced correlation satellite at 12.8 eV compared with the self-convoluted Cu 3d pDOS (SM : Fig. S3). The energy separation between the main peaks gives U$_{dd}$ = 6.5 $\pm$0.5 eV.}}
\end{figure}

\begin{figure}
\centering
\includegraphics[width=\columnwidth]{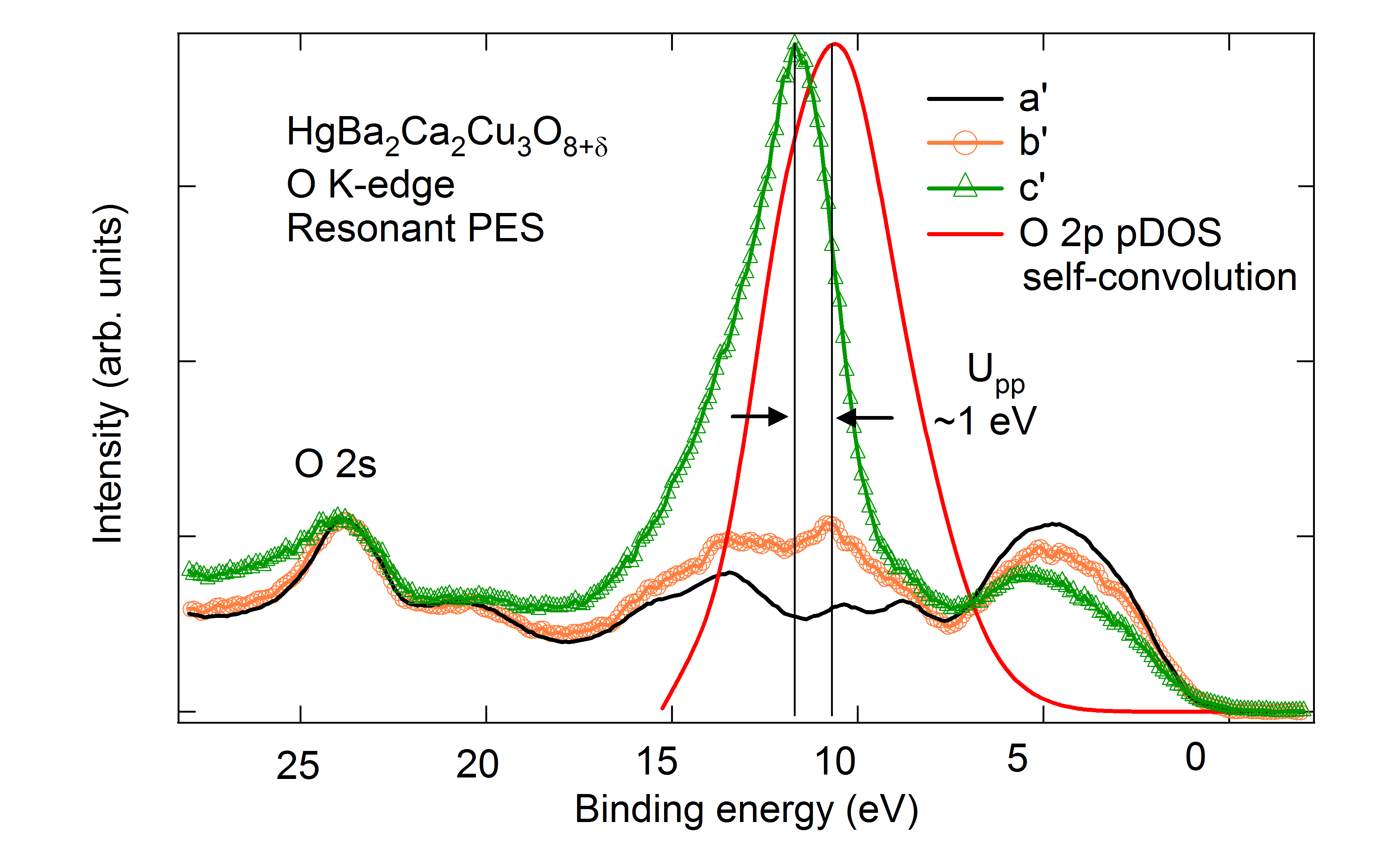}
\caption{\label{las} {(color online)(A) The O 1s-2p resonant PES spectra (labelled $\bf{a'-c'}$, from Fig. S4) of Hg1223 measured at T = 20 K (normalized at the O 2s feature at $\sim$ 24 eV BE) showing the resonantly enhanced correlation satellite at 11.6 eV compared with the self-convoluted O 2p pDOS. The energy separation between the main peaks gives U$_{pp}$ = 1.0 $\pm$0.5 eV.}}\end{figure}

In order to estimate the on-site U$_{dd}$, we identify the dominantly Cu 3d pDOS in the valence band. We do this by subtracting out the normalized ultraviolet (h$\nu$ = 21.2 eV) valence band spectrum ( which consists mainly of the O 2p pDOS ) from the soft x-ray (h$\nu$ = 1200 eV) spectrum, as shown in SM Fig.S4. We carry out a self-convolution of the Cu 3d pDOS and compare it with the spectrum showing the resonantly enhanced Cu 3d correlation satellite which occurs at 12.8 eV BE(Fig. 3). The spectra labelled $\bf{a}$ and $\bf{b}$ are the same as shown in Fig. 2(B), but drawn on an expanded y-scale and normalised at the O 2s core level peak at $\sim$ 24 eV BE. The energy separation between the main peaks of the resonantly enhanced spectrum and the two-hole spectrum gives a measure of on-site U$_{dd}$ = 6.5 $\pm$0.5 eV, consistent with earlier work on other cuprates.\cite{Marel,Balzarotti,Tjeng,BarDeroma}

Next, we carry out a similar analysis to determine the on-site U$_{pp}$, by measuring the O K-edge XAS and O 1s-2p resonant valence band PES (data shown in SM Fig. S5). The O 1s-2p resonant PES identifies the O KVV Auger feature originating in the O 2p correlation satellite, occuring at $\sim$11.6 eV. In Fig. 4, we compare the self convolution of the O 2p pDOS with the spectra $\bf{a'-c'}$ showing the resonantly enhanced correlation satellite. The spectra labelled $\bf{a'-c'}$ are the same as shown in Fig. S4, but plotted on an expanded y-scale and normalised at the O 2s peak at $\sim$ 24 eV BE. From the energy separation between the main peaks, we estimate an on-site U$_{pp}$ = 1.0 $\pm$0.5 eV.
Such a low value of U$_{pp}$ for 3d transition metal oxides, in general, and for hole-doped cuprates in particular, has not been reported to date.
Since a large U$_{dd}$ coexisting with a small U$_{pp}$ is unusual, we checked its validity using another method.

\begin{figure}
\centering
\includegraphics[width=\columnwidth]{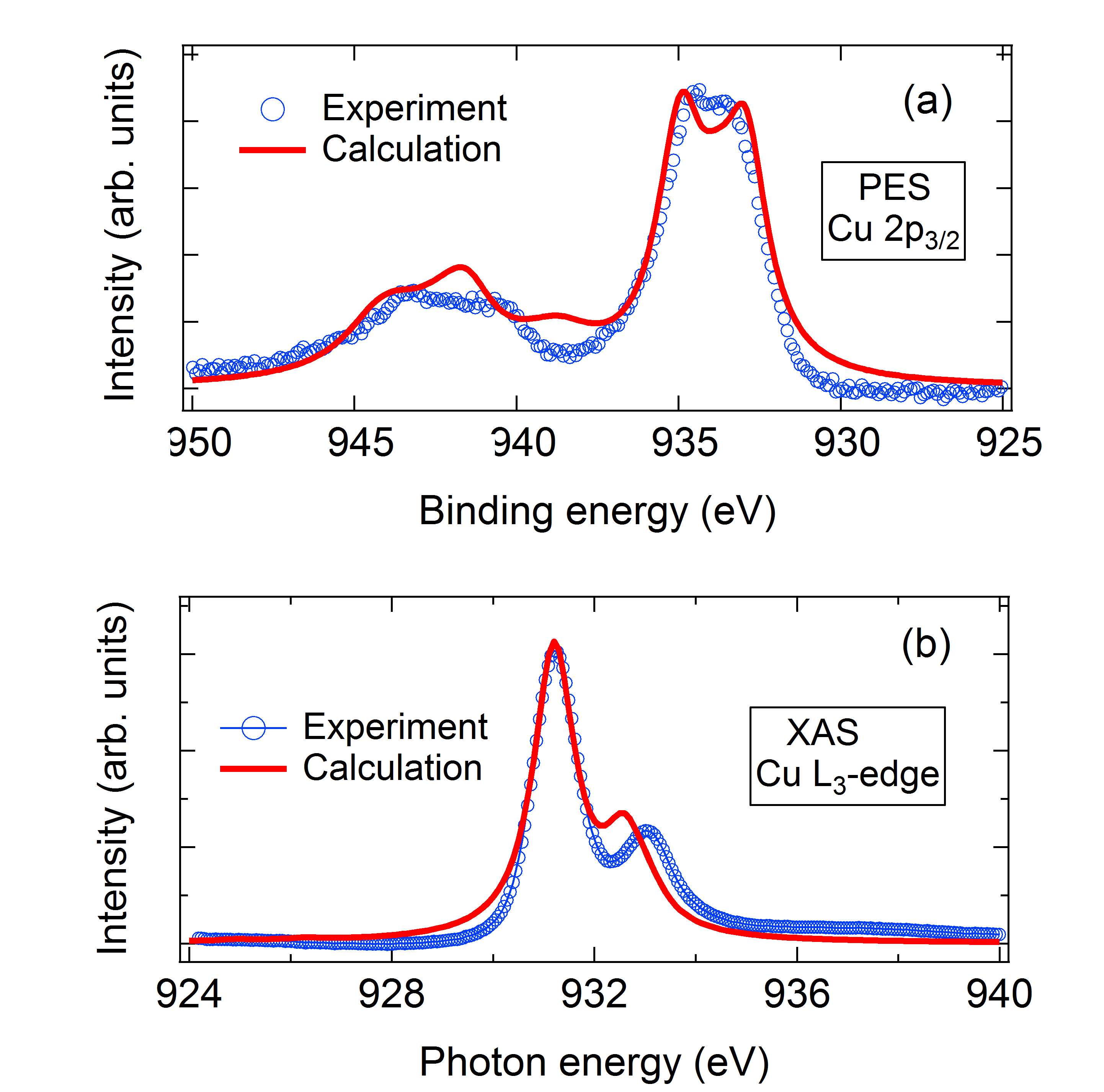}
\caption{\label{las} {(color online)(a) The Cu 2p$_{3/2}$ PES of Hg1223 measured at T = 20 K compared with a Cu$_2$O$_7$ cluster calculation with non-local screening.
(b) The Cu L$_3$-edge XAS of Hg1223 measured at T = 20 K compared with a Cu$_2$O$_7$ cluster calculation with non-local screening.}}
\end{figure}

It is well-known that the Cu 2p core-level PES is best explained by
model many-body cluster calculations which go beyond a single CuO$_{4}$ plaquette by
including non-local screening from a neighbouring CuO$_{4}$ plaquette.\cite{Veenendaal,Okada}  
We have carried out such calculations for a Cu$_{2}$O$_{7}$ cluster and compared the results with the experimentally measured Cu 2p PES. While non-local calculations for Cu L-edge XAS have shown an additional satellite feature,\cite{Veenendaal} it has not been directly compared with experiment as most cuprates show only a single peak XAS spectrum which gets broadened with doping\cite{Tjeng,CTChen,Nucker}. However, the Hg-based cuprates have shown a weak well-separated satellite feature.\cite{Pellegrin} In the present study, we also see a clear satellite which is reproduced by calculations. The calculations (see SM for details) were carried out for the simplest case with 3 holes in the Cu$_{2}$O$_{7}$ cluster, corresponding to 50\% hole doping. 

Fig. 5(a) shows the Cu 2p$_{3/2}$ core level PES spectrum plotted with the calculated spectrum. The experimental data shows a main peak consisting of two features at 933.4 eV and 934.5 eV, and broad weak intensity satellite structures between $\sim$939 eV to $\sim$946 eV. The calculated spectrum(Fig. 5(a)) was obtained using $\Delta$ = 1.3 eV, where $\Delta$ is the charge-transfer energy,  the O 2p-Cu 3d overlap integral t$_{pd}$ = 0.45 eV, the O 2p-O 2p overlap integral t$_{pp}$ = 0.3 eV, U$_{dd}$ = 6.5 eV, U$_{pp}$ = 1.0 eV and the Coulomb interaction due to the core hole U$_{dc}$ = 8.5 eV. While the calculations reproduce all the features of the experimental data but with small discrepancies for the satellite features, we use the above parameters as they give the best fit to the Cu 2p PES and L-edge XAS  spectra simultaneously. We checked that the well-screened feature at 933.4 eV in the Cu 2p$_{3/2}$ main peak (the lowest BE feature) is dominated by the Cu1 $\underline{c}$d$^{10}\underline{L^1}$ : Cu2 d$^{9}\underline{L^1}$ configuration (the prefix Cu1 and Cu2 identify the two Cu sites and the core hole $\underline{c}$ is created on the Cu1 site, see Fig. S6 in SM). Thus, the well-screened feature involves `non-local' screening from the Cu2 site CuO$_{4}$ plaquette, leading to an effective ZRS state consistent with earlier work.\cite{Veenendaal, Okada} The feature at 934.5 eV in the main peak is attributed to `local' screening from the Cu1 site plaquette as it is dominated by Cu1 $\underline{c}$d$^{9}\underline{L^1}$ : Cu2 d$^{9}$ and Cu1 $\underline{c}$d$^{10}\underline{L^2}$ : Cu2 d$^{9}$ states, with admixture from Cu1 $\underline{c}$d$^{10}\underline{L^2}$ : Cu2 d$^{10}\underline{L^1}$.
The satellite features are dominated by ionic configurations on the Cu1 site namely Cu1 $\underline{c}$d$^{9}$ : Cu2 $\underline{c}$d$^{9}\underline{L^1}$ and Cu1 $\underline{c}$d$^{9}$ : Cu2 $\underline{c}$d$^{10}\underline{L^2}$ with admixture from $\underline{c}$d$^{9}\underline{L^1}$ : Cu2 d$^{9}$ and $\underline{c}$d$^{10}\underline{L^1}$ : Cu2 d$^{8}$ configurations. 

Similarly, Fig. 5(b) shows the Cu L$_{3}$ XAS spectrum compared to calculations carried out with the same  paramaters as for the Cu 2p core level PES discussed above. The main peak in Cu L$_{3}$ XAS spectrum at $\sim$931 eV is dominated by the $\underline{c}$d$^{9+1}\underline{L^1}$ : Cu2 d$^{10}\underline{L^1}$ state involving `non-local' screening from the Cu2 site, with admixture from $\underline{c}$d$^{9+1}\underline{L^1}$ : Cu2 d$^{9}$ state. The satellite feature at $\sim$933 eV is dominated by $\underline{c}$d$^{9+1}\underline{L^1}$ : Cu2 d$^{9}$ state with admixture from $\underline{c}$d$^{8+1}$ : Cu2 d$^{10}$$\underline{L^1}$. 
We carried out additional checks (see SM, Figs. S7 and S8) for the validity of the parameters used for the calculations. The calculations indicate deviations from experimental data if U$_{dd}$ $<$ 4 eV or U$_{pp}$ $\geq$ 3 eV. This confirms a large U$_{dd}$ coexisting with a small U$_{pp}$ in the highest-T$_{c}$ Hg1223 cuprate, and indicates an electronic inhomogeneity on the length scale of a CuO$_4$ plaquette. 

Our results thus indicate the necessity of distinguishing between U$_{dd}$ and U$_{pp}$, which originates from dynamical screening effects at the Cu- and O-site.\cite{Werner} It is known that the strength of correlations depends on the presence or absence of apical oxygens, with stronger correlations in hole-doped LSCO having apical oxygen, and weaker correlations in electron-doped NCCO, having no apical oxygen.\cite{Weber} For the single-band Hubbard model, a very recent extension showed that U$_{dd}$ is proportional to the inverse of the bond distance between apical oxygen and copper atoms even for the Hg-based cuprates.\cite{Jang} Since Hg1223 has 3 Cu-O layers : the inner layer has no apical oxygen and the two outer layers has pyramidal CuO$_5$ co-ordination, the effective correlations are weakest in the Hg1223 system quite like the electron-doped cuprates.\cite{Jang} In addition, in the only study of a three band model expicitly including U$_{dd}$ and U$_{pp}$, it was shown that the static U$_{dd}$ = 7.0 eV and U$_{pp}$ = 4.64 eV for La$_2$CuO$_4$.\cite{Werner} The authors also discussed that the frequency dependent local self-energy is larger for the Cu site with 4 nearest neighbour oxygens, and smaller for the O site with 2 nearest neighbour Cu atoms. Finally,
since the charge carriers in the hole-doped cuprates have large O 2p hole character, coexistence of a small U$_{pp}$ and large U$_{dd}$ suggests pairing would be favored for the weakly correlated oxygen holes in the highest-T$_{c}$ Hg1223 cuprate.

In conclusion, resonant valence band PES across the O K-edge and Cu L-edge  identify correlation satellites due to two-hole final states in Hg1223. Analyses using the measured O 2p and Cu 3d partial DOS show that on-site Coulomb energy for the O-site (U$_{pp}$ =~1.0$\pm$0.5 eV) is much smaller than that for Cu-site (U$_{dd} =~$6.5$\pm$0.5 eV). Cu$_{2}$O$_{7}$-cluster calculations with non-local screening for the Cu 2p core level PES and Cu L-edge XAS spectra are used to confirm the U$_{dd}$ and U$_{pp}$ values, and provide evidence for the Zhang-Rice singlet state in optimally doped Hg1223. In contrast to known results of U$_{pp}$ $\sim$ U$_{dd}$ for other hole-doped cuprates as well as 3d-transition metal oxides in general, the present results indicate weakly correlated oxygen holes in the highest-T$_{c}$ cuprate Hg1223. 

Acknowledgements : AC thanks the Institut Jean Lamour, Universit\'{e} de Lorraine for hospitality and financial support during the course of this work. The synchrotron radiation
experiments were performed at BL17SU,
SPring-8, with the approval of RIKEN (Proposal
No. 20140019).



\pagebreak
\onecolumngrid

\begin{center}
\textbf{\large 
{Supplementary Material : Evidence for weakly correlated oxygen holes in the highest-T$_{c}$ cuprate superconductor HgBa$_2$Ca$_2$Cu$_3$O$_{8+\delta}$}}

\vspace{0.8cm}
A. Chainani, M. Sicot, Y. Fagot-Revurat, G. Vasseur, J. Granet, B. Kierren, L. Moreau, M. Oura, A. Yamamoto,

Y. Tokura and D. Malterre.
\end{center}
\vspace{0.8cm}

\setcounter{equation}{0}
\setcounter{figure}{0}
\setcounter{table}{0}
\setcounter{page}{1}
\makeatletter
\renewcommand{\theequation}{S\arabic{equation}}
\renewcommand{\thefigure}{S\arabic{figure}}
\renewcommand{\bibnumfmt}[1]{[S#1]}
\renewcommand{\citenumfont}[1]{S#1}

\twocolumngrid
\subsection{Experimental details of the sample preparation, characterization and the spectroscopy measurements:}

The polycrystalline Hg1223 sample was prepared from a mixture of HgO and a
precursor. The precursor was  prepared from BaCO$_{3}$,
CaCO$_{3}$, and CuO at 950 C
in flowing high-purity oxygen gas. It was found necessary to use
a slightly Hg-deficient
starting composition of Hg$_{0.75}$Ba$_{2}$Ca$_{2}$Cu$_{3}$O$_{y}$ (HgO : Ba$_{2}$Ca$_{2}$Cu$_{3}$O$_{y}$ = 0.75 : 1) to avoid impurity phases.\cite{Yamamoto2} The mixture was pressed into a cylindrical
pellet, charged into a gold capsule, and then heated at
830 C under 2 GPa for 30 min, using a cubic-anvil high-pressure
apparatus (TRY Engineering, 180-ton press).
The powder
X-ray diffraction patterns confirmed the single phase of
Hg1223. Fig. S1 shows the superconducting transition with a $T_{c,onset}$ = 134.4 K and $T_{c,0}$ = 133.2 K, from electrical resistivity measurements (PPMS, Quantum
Design). The inset shows the wide range resistivity data upto 300 K showing the nearly linear resistivity 
as a function of temperature, typical of high-T$_c$ cuprate superconductors.

\begin{figure}[h!]
\includegraphics[width=\columnwidth]{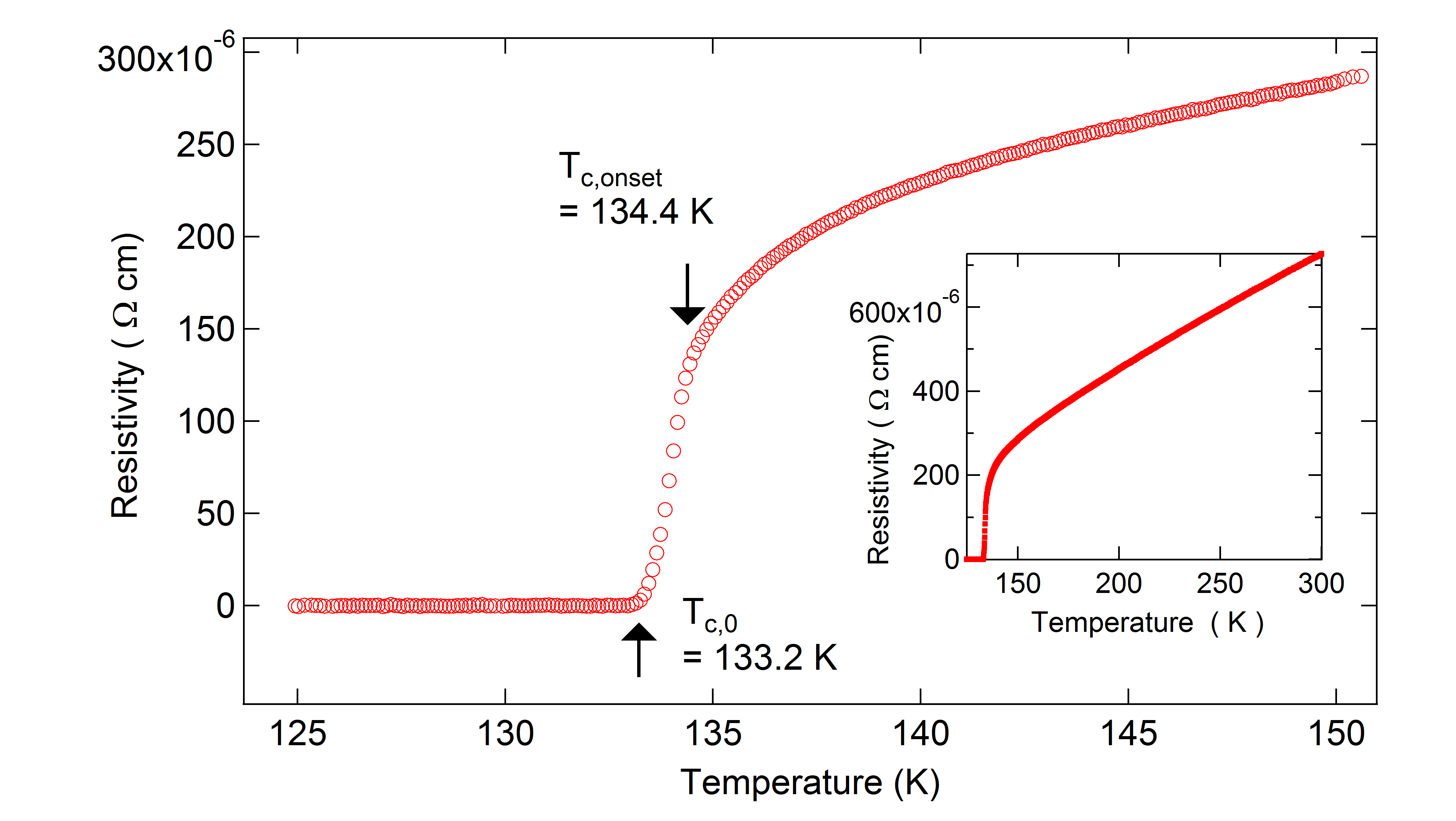}
\caption{(color online) Electrical resistivity of HgBa$_2$Ca$_2$Cu$_3$O$_{8+\delta}$ measured as a function of temperature showing the superconducting transition with a $T_{c,onset}$ = 134.4 K and $T_{c,0}$ = 133.2 K. The inset shows the wide range resistivity data upto 300 K.}\label{S1}
\end{figure}

Ultraviolet (UV)-PES using a He I discharge lamp was carried out
at University of Lorraine. The spectrometer is equipped with a 
Gammadata-Scienta SES2002 hemispherical electron analyzer. The sample was cooled using a
flowing liquid nitrogen cryostat and the sample temperature was T = 120 K. 
The total energy resolution was 
40 meV for the UV-PES measurements.  Soft x-ray PES
measurements were carried out at beamline BL17SU,
SPring-8 using incident photon energies 
of h$\nu$ = $\sim$524 eV-1200 eV and a spectrometer equipped with a Gammadata-Scienta SES2002 hemispherical electron analyzer. The total energy resolution was 
about 160 - 280 meV for the SX-PES measurements. 
XAS measurements
were recorded in the total electron yield mode. SX-PES measurements were carried out at a sample temperature of
T = 20 K obtained using a flowing liquid He cryostat.
The measurements were carried out in a vacuum below
4 x 10$^{-8}$ Pa and clean sample surfaces were obtained by
cleaving the sample. For the UV-PES measurements, the cleaving
was done in the preparation chamber at room temperature 
and the sample was transferred immediately into the
analysis chamber and cooled, while for the SX-PES, XAS
and Resonant-PES measurments, the cleaving was carried out
in-situ at T = 20 K. The spectra were calibrated using
the Fermi level (E$_{F}$ ) measured from a gold film evaporated onto the sample holder.  
The Resonant-PES measurements were also calibrated by measuring the gold spectrum at the start and end of every cycle of photon energies used for the Cu L-edge and O K-edge measurements.

\subsection{Core level PES and wide range valence band PES measurements:}

\begin{figure}[h!]
\includegraphics[width=\columnwidth]{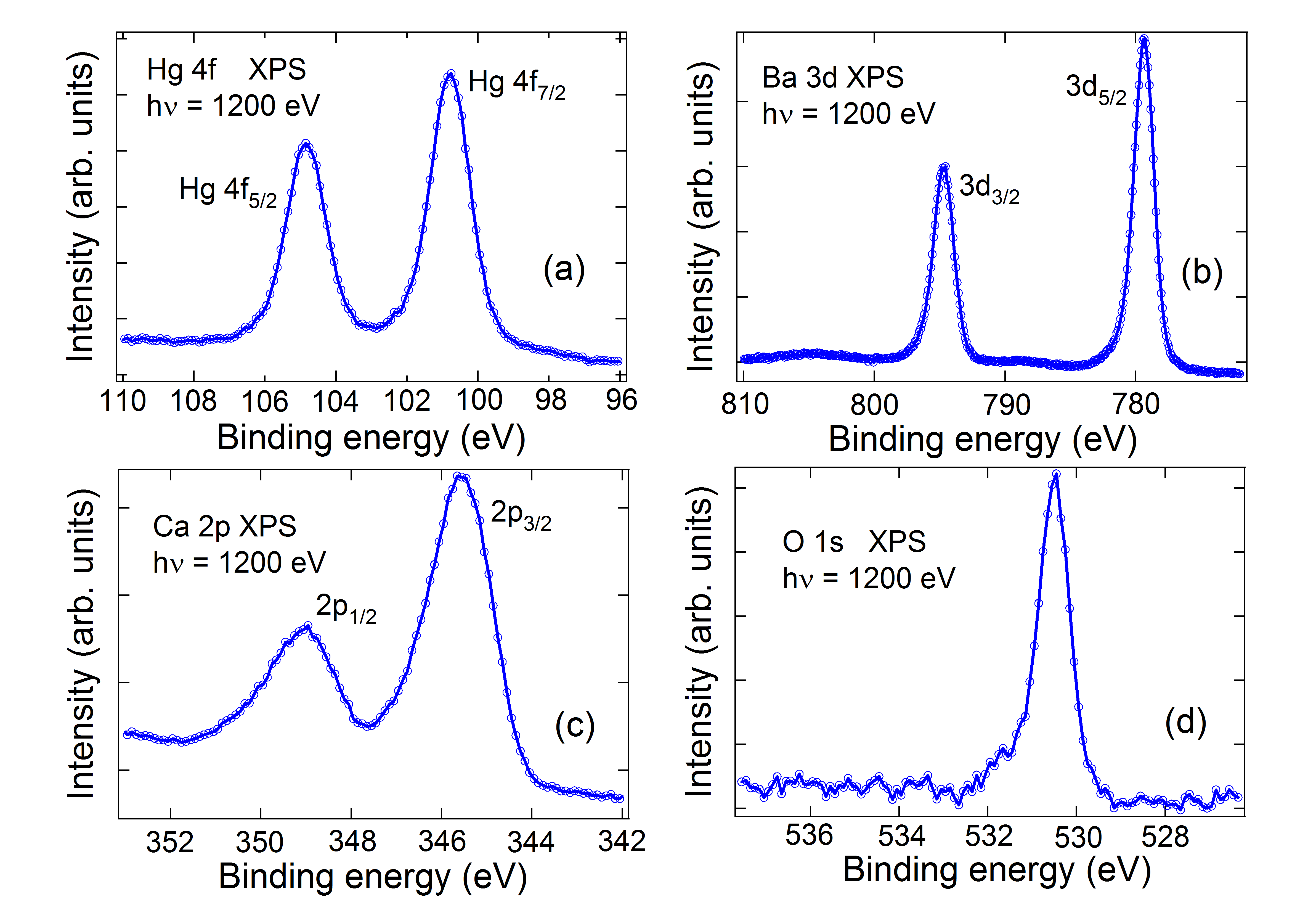}
\caption{ (color online) The soft x-ray core level photoemission spectra of (a) Hg 4f, (b) Ba 3d , (c) Ca 2p and (d) O 1s of HgBa$_2$Ca$_2$Cu$_3$O$_{8+\delta}$, measured at T = 20 K using an incident photon energy of
h$\nu$ = 1200 eV.}\label{S2}
\end{figure}

Fig. S2(a-d) shows the core levels of Hg 4f, Ba 3d , Ca 2p and O 1s, respectively, measured at T = 20 K using a photon energy h$\nu$ = 1200 eV. The Hg 4f (Fig. S2(a)) spectrum exhibits clean single peaks for the Hg 4f$_{7/2}$ and 4f$_{5/2}$ spin-orbit split levels, occuring at 100.8 eV and 104.8 eV. These values are typical of Hg$^{2+}$ states as in HgO.\cite{NIST} This result rules out the presence of Hg$^{3+}$, and hence mixed valency of Hg ions in Hg1223, which was claimed in an early study based on Hg 4f levels showing doublet features for the 4f$_{7/2}$ and 4f$_{5/2}$ levels.\cite{Gopinath} We believe the difference between the present study and earlier work stems from the fact that we have used a high pressure synthesis and we obtain optimally doped samples with a T$_{c}$ = 134 K, which corresponds to a hole doping p = 0.2.\cite{Yamamoto2} In contrast, the earlier study, which used samples made by a solid state reaction method, had inferred their sample to be underdoped based on an analysis that Hg$^{3+}$ would effectively remove holes.\cite{Gopinath} Fig. S2(b) shows the Ba 3d$_{5/2}$ and 3d$_{3/2}$ core levels as sharp single peaks at 779.4 eV and 794.7 eV, respectively. These values are typical of Ba$^{2+}$ as in BaO.\cite{NIST} Similarly, Fig. S2(c) shows the Ca 2p$_{3/2}$ and 2p$_{1/2}$ core levels as single peaks at 345.5 eV and 349.0 eV, and are typical of Ca$^{2+}$ states.\cite{NIST} Fig. S2(d) shows the oxygen 1s core level with an essentially single peak at 530.5 eV, although a very weak feature at 531.8 eV can also be discerned. The high binding energy weak feature at 531.8 eV is attributed to suface non-stoichiometry, which is seen even for cleaved single crystal metal oxides.\cite{Parmigiani}

\begin{figure}[h!]
\includegraphics[width=\columnwidth]{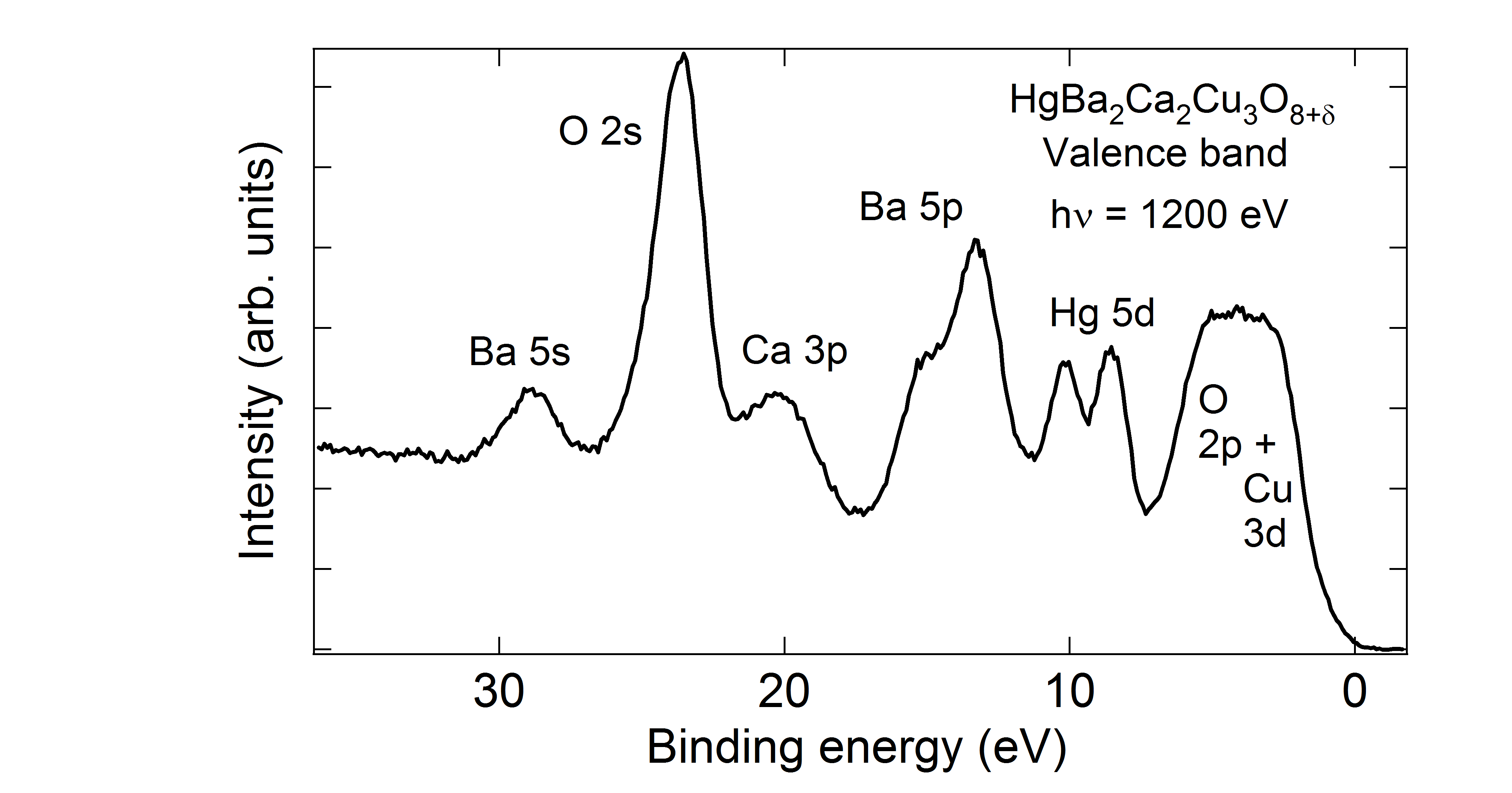}
\caption{ The wide energy range valence band and shallow core levels of HgBa$_2$Ca$_2$Cu$_3$O$_{8+\delta}$ measured at T = 20 K.}\label{S2}
\end{figure}

\begin{figure}
\centering
\includegraphics[width=\columnwidth]{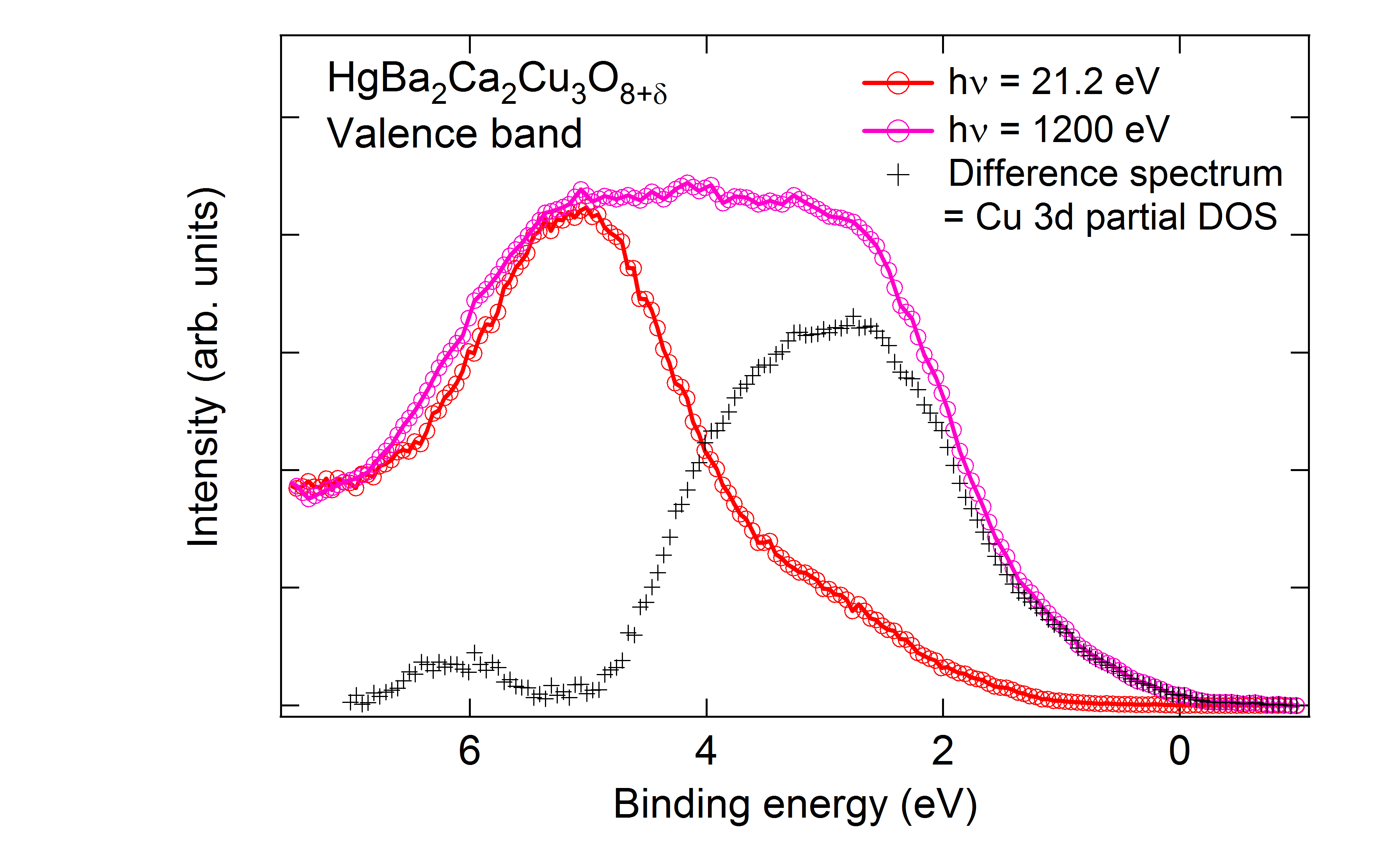}
\caption{\label{las} { (color online) (A) The Cu 3d pDOS of HgBa$_2$Ca$_2$Cu$_3$O$_{8+\delta}$ was experimentally determined by comparing the valence band spectra measured using h$\nu$ = 21.2 eV and h$\nu$ = 1200 eV. The spectra are normalised at 5.0 eV binding energy corresponding to the feature assigned to dominantly O 2p character states. The difference spectrum between h$\nu$ = 21.2 eV and h$\nu$ = 1200 eV spectra provides a measure of the dominantly Cu 3d pDOS.}}
\end{figure}

\begin{figure}
\centering
\includegraphics[width=\columnwidth]{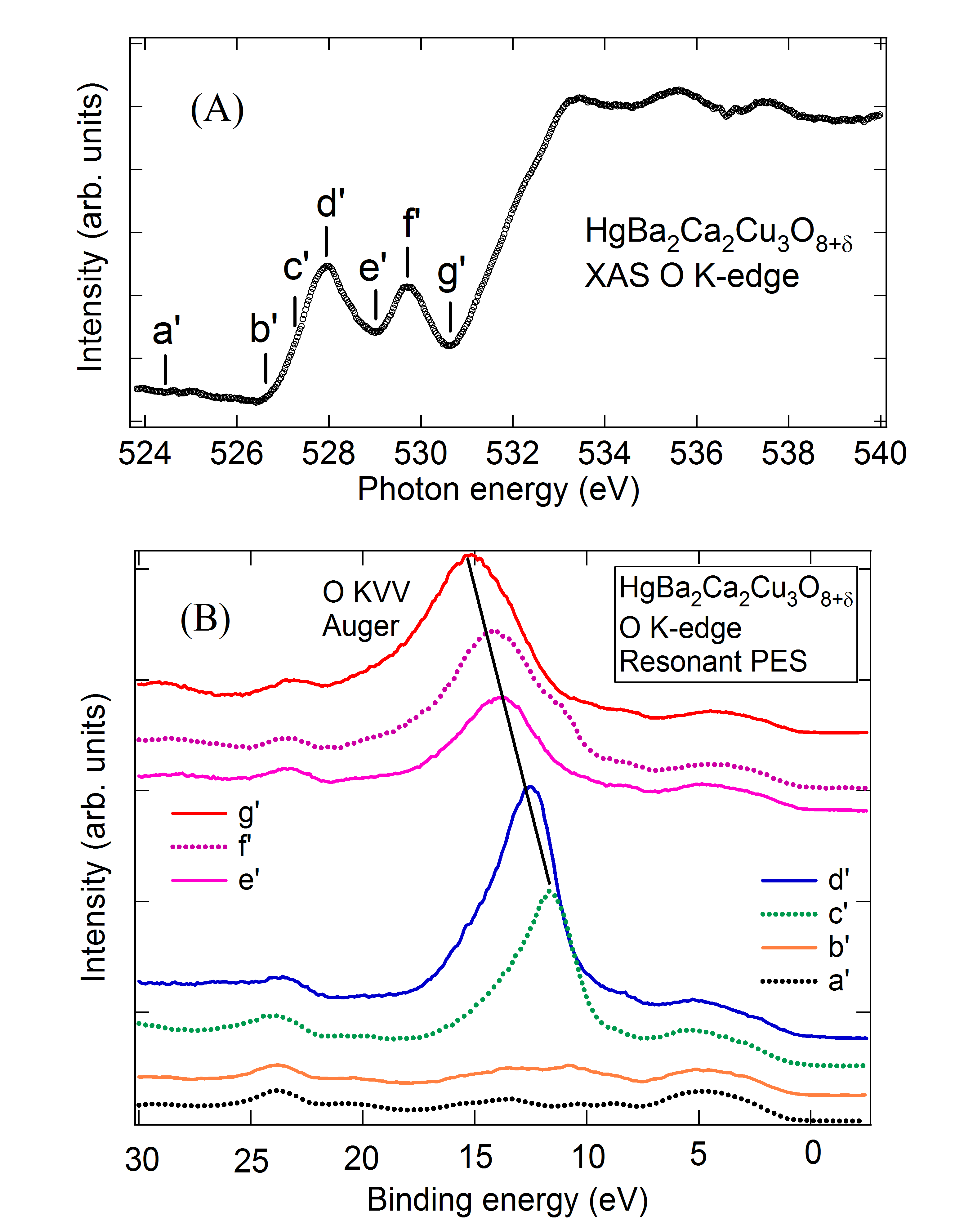}
\caption{\label{las} { (color online) (A) The O K-edge(1s-2p) X-ray absorption spectrum of HgBa$_2$Ca$_2$Cu$_3$O$_{8+\delta}$ measured at T = 20 K. The labels $\bf{a'-g'}$ mark the photon energies used for measuring the O 1s-2p resonant valence band photoemission spectra of HgBa$_2$Ca$_2$Cu$_3$O$_{8+\delta}$ shown in panel (B). 
The resonant photoemission spectra are normalized for the incident photon flux and number of scans. The 2p$^4$ resonance satellite arising due to on-site U$_{pp}$ correlations gets enhanced but stays fixed upto a photon energy of 527.6 eV (label $\bf{c'}$ in panel A) and evolves into the O K$_{3}$VV Auger satellite for higher photon energies.}}
\end{figure}

Fig. S3 shows the wide range valence band spectrum of Hg1223 measured at T = 20 K using h$\nu$= 1200 eV.
The spectrum shows the shallow core levels of Ba 5s at about 29 eV, the O 2s peak at 23.5 eV, and the Ca 3p at 20 eV binding energy. The Ba 5p$_{3/2}$ and 5p$_{1/2}$ doublet is observed at 13.3 eV and 14.8 eV, respectively, while the Hg 5d$_{5/2}$ and 5d$_{3/2}$ levels are seen at 8.5 eV and 10.1 eV binding energies. The valence band consisting of the Cu 3d and O 2p states occurring between the Fermi level and about 6 eV binding energy are discussed in detail in Fig. 1 of the main paper, which compares the valence band spectra obtained using incident photon energies of h$\nu$ = 21.2 eV, 524.4 eV, 926.3 eV and 1200 eV. Based on atomic cross-sections as a function of incident photon energy, the photon energy dependence indicates that the 5 eV feature seen in the h$\nu$ = 21.2 eV spectrum is dominated by the O 2p partial density of states (pDOS). This is consistent with band structure calculations of Hg1223.\cite{Rodriguez2}
In order to separate out the O 2p and Cu 3d states, in Fig. S4 we plot the spectrum measured using h$\nu$ = 21.2 eV with the h$\nu$ = 1200 eV spectrum.
The spectra are normalized at 5 eV binding energy corresponding to the feature due to dominantly O 2p states. The difference spectrum obtained by subtracting the h$\nu$ = 21.2 eV from the h$\nu$ = 1200 eV spectrum is representative of the dominantly Cu 3d pDOS(Fig. S4). A numerical self-convolution of the obtained Cu 3d pDOS is carried out to obtain the two hole spectrum and compared with the spectrum showing the Cu 3d correlation satellite, as shown in Fig. 3 of the main paper.

\subsection{Cini-Sawatzky method for estimating on-site Coulomb energy:}

Early work by Cini\cite{Cini} and Sawatzky\cite{Sawatzky} showed that on-site Coulomb energies can be quantified and the method has been applied to several TM oxides.\cite{Marel,Balzarotti,Tjeng,BarDeroma,Ishida,Post,Park,AC,DD,Ghijsen} The method is : determine the one-electron removal (single-hole) valence band (VB) partial density of states(pDOS), numerically evaluate the two-valence-hole energies as a self-convolution of the single-hole states, and then compare with the measured correlation satellite associated with the two-valence-hole (VV) Auger final state for estimating the site-specific Coulomb energies in compounds. We follow the same procedure for obtaining U$_{dd}$ and U$_{pp}$ via measurements of the O 2p and Cu 3d VB pDOS, and the O KVV and Cu LVV Auger spectra.

\subsection{O K-edge XAS and O 1s-2p resonant valence band PES measurements:}

In Fig. S5(A), we show the O K-edge XAS which exhibits a two peak structure at $\sim$528 eV and $\sim$530 eV photon energies (labelled $\bf{d'}$ and $\bf{f'}$
). These two peaks are due to O 2p states hybridized with Cu 3d states and are attributed to the doped hole states and the upper Hubbard band, respectively.\cite{CT,Fink} The higher lying features at $\sim$533 eV-536 are due to the Hg-Ba-Ca p-character states. We then carried out O 1s-2p resonant valence band PES at the photon energies labelled $\bf{a'-g'}$ and the spectra are correspondingly labelled and shown in Fig. S5(B). The off-resonance spectrum $\bf{a'}$ is the same as the h$\nu$ = 524.4 eV spectrum of Fig. 1. As we increase the photon energies from $\bf{a'}$ to $\bf{g'}$, we see a systematic evolution of the valence band spectra. In particular, at the energy $\bf{c'}$ = 527.6 eV, we see a clear high intensity feature centered at 11.6 eV BE. On increasing the photon energy further, we see a shift of this feature to higher binding energies, with the magnitude of the energy shift tracking the increase in photon energy. This identifies the resonance feature as the O KVV Auger state, with two final state holes(VV) in the valence band. This confirms its origin to be the O KVV correlation satellite and we analyse it further to obtain an on-site U$_{pp}$ = 1.0 $\pm$0.5 eV, as discussed in the main text(Fig. 4).

\subsection{Model many-body calculations including non-local screening for a Cu$_{2}$O$_{7}$ cluster 
:}

We have carried out model many-body Hamiltonian calculations for the Cu 2p XPS and Cu L-edge XAS using a Cu$_2$O$_7$ cluster with non-local screening, following the work of Veenendaal et al\cite{Veenendaal3}. We consider the simplest case of hole doping, namely, one extra hole in the Cu$_2$O$_7$ cluster where the undoped reference state has one hole per Cu site. Thus, we have three holes in the Cu$_2$O$_7$ cluster corresponding to a hole-doping content of 50\%.
We use the notation for the Cu$_2$O$_7$ cluster(see Fig. S6) with the Cu site in the left CuO$_4$ plaquette
labelled as Cu1 and that in the right CuO$_4$ plaquette is labelled as Cu2.

\begin{figure}
\centering
\includegraphics[width=\columnwidth]{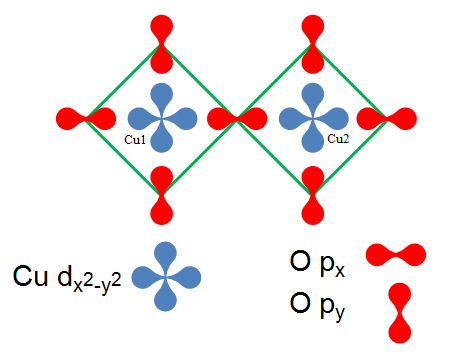}
\caption{\label{las} {(color online) The Cu$_2$O$_7$ cluster used for the non-local screening calculations, with the Cu site in the left CuO$_4$ plaquette
being Cu1 and the one in the right CuO$_4$ plaquette is Cu2. We introduce 3 holes in the cluster to represent 50\% hole doping in the Cu$_2$O$_7$ cluster.}}
\end{figure}

For the assumed case of the core hole on the Cu1 site, the purely ionic character state is given by 
$|$ Cu1 d$^8$ ; Cu2 d$^9$ $>$ and its complementary state is given by  
the $|$ Cu1 d$^8$ ; Cu2 d$^9$ $>$ state.
For the d$^8$ ionic configuration, we consider d$^{9}$$\underline{L^1}$ and d$^{10}$$\underline{L^2}$
charge-transferred states,
while for the d$^9$ configuration, we consider d$^{10}$$\underline{L^1}$ charge-transferred state.
The electronic parameters entering the calculation are $\Delta$, the charge-transfer energy,  the O 2p-Cu 3d overlap integral t$_{pd}$, the O 2p-O 2p overlap integral t$_{pp}$, U$_{dd}$, U$_{pp}$ and the Coulomb interaction due to the core hole U$_{dc}$. Here, we have restricted U$_{dd}$ and U$_{pp}$ to the experimentally estimated values. We have neglected multiplet effects in the calculations.
The pure `local screening' state is one in which the ligand from the Cu1 site plaquette with the core hole takes part in screening e.g. $\underline{c}$d$^{n+1}$$\underline{L^1}$ ; Cu2 d$^{m}$, where n and m correspond to the initial ionic states. The pure `non-local screening' corresponds to the case when the core hole is on the Cu1 site while the ligand from the Cu2 site plaquette participates in the screening process e.g. Cu1 $\underline{c}$d$^{n+1}$ ; Cu2 d$^{m}$$\underline{L^1}$. And we can also have states like Cu1 $\underline{c}$d$^{n+1}$$\underline{L^1}$ ; Cu2 d$^{m+1}$$\underline{L^1}$, or Cu1 $\underline{c}$d$^{n+2}$$\underline{L^1}$ ; Cu2 d$^{m}$$\underline{L^1}$, etc. which have local and non-local screening. 
The non-local screening indicates that the more stable configuration requires the ligand hole screening on the Cu2 site and represents the Zhang-Rice singlet (ZRS) state.\cite{Veenendaal3} As discussed in the introduction, it has been shown that the ZRS state survives in the doped hole case using core level PES\cite{Veenendaal942,Taguchi2} and spin-polarized resonant photoemission measurements.\cite{Brookes2}
In spite of the restricted number of basis states, we found that these
basis states were sufficient to reproduce all the features in the experimental 
Cu 2p$_{3/2}$ PES and Cu L$_{3}$-edge XAS spectra as shown in Fig. 5 and discussed in the main paper. 

In order to further check the validity of the estimated U$_{dd}$ and U$_{pp}$ values, we have carried out an extensive check by varying U$_{dd}$ and U$_{pp}$ and keeping all other parameters fixed. 
The results are shown in Figs. S7 and S8.

\begin{figure}
\centering
\includegraphics[width=\columnwidth]{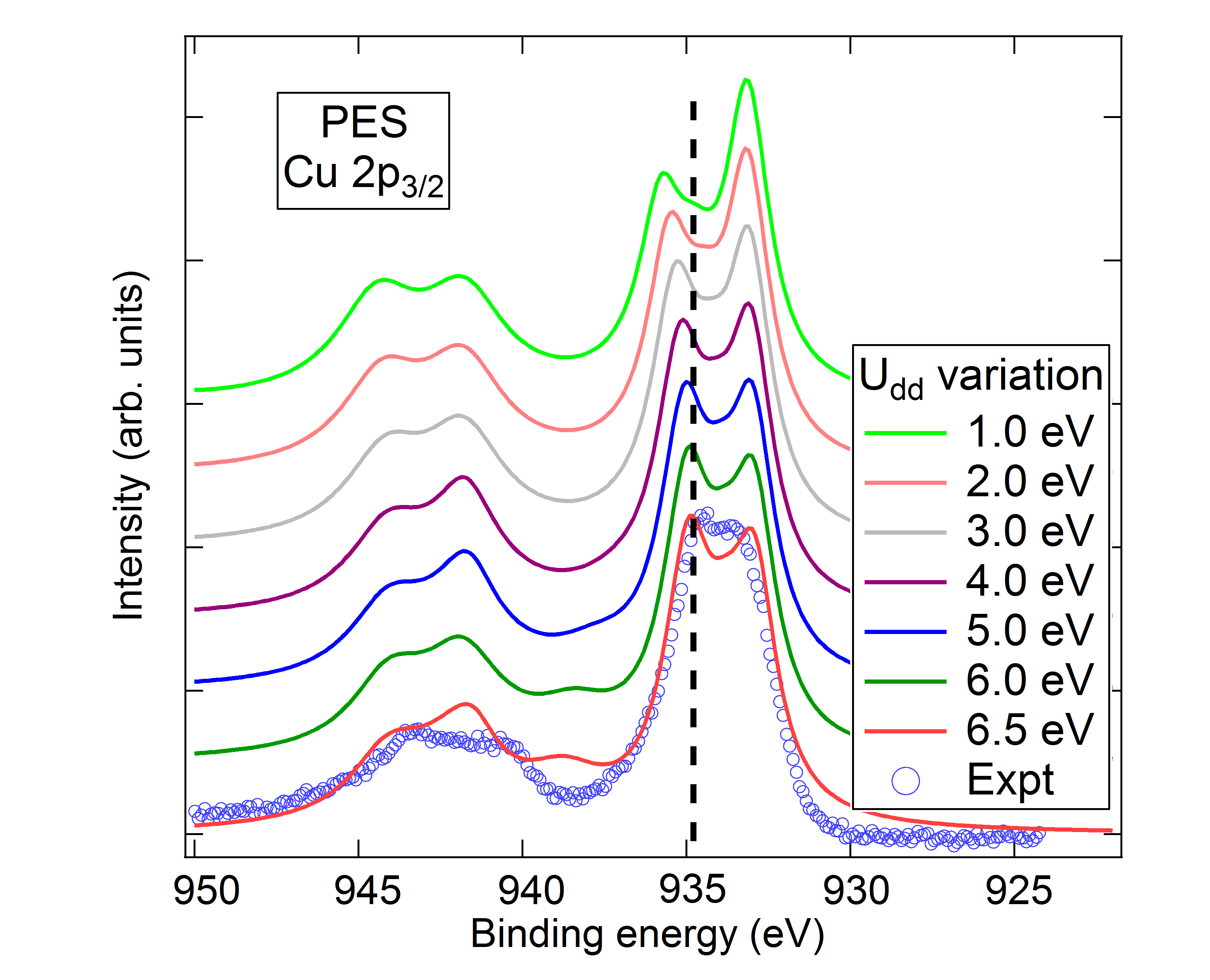}
\caption{\label{las} {(color online) The variation of the calculated Cu 2p$_{3/2}$ X-ray photoemission spectra as a function of the on-site U$_{dd}$. The calculations deviate from the experimental 
Cu 2p spectrum of HgBa$_2$Ca$_2$Cu$_3$O$_{8+\delta}$ for U$_{dd}$ $<$ 4 eV.}}
\end{figure}

\begin{figure}
\centering
\includegraphics[width=\columnwidth]{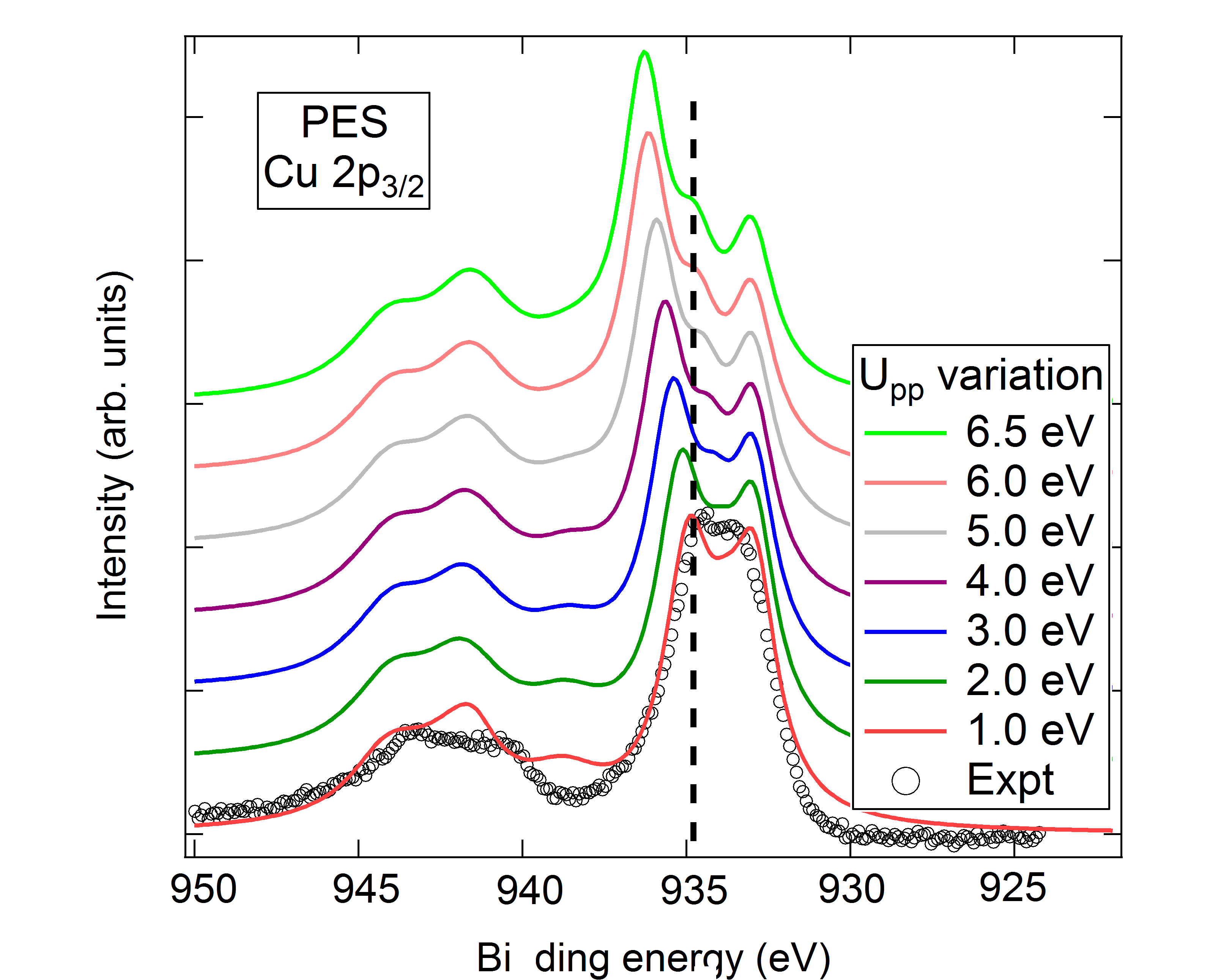}
\caption{\label{las} {(color online) The variation of the calculated Cu 2p X-ray photoemission spectra as a function of the on-site U$_{pp}$. The calculations deviate from the experimental 
Cu 2p spectrum of HgBa$_2$Ca$_2$Cu$_3$O$_{8+\delta}$ for U$_{pp}$ $\geq$ 3 eV.}}
\end{figure}

Fig. S7 shows the variation of the calculated Cu 2p$_{3/2}$ X-ray photoemission spectra as a function of the on-site U$_{dd}$, which was varied from the optimal value of U$_{dd}$ = 6.5 eV down to U$_{dd}$ = 1.0 eV, keeping all other parameters fixed. The calculations do not show much change down to a U$_{dd}$ value of 4 eV, but on reducing U$_{dd}$ further, the calculated spectra start showing deviations from the experimental 
Cu 2p spectrum of HgBa$_2$Ca$_2$Cu$_3$O$_{8+\delta}$. In particular, for U$_{dd}$ $<$ 4 eV, the energy separation between the main peaks consisting of the dominantly local (BE = 934.5 eV)  and non-local (BE (933.4 eV) screened states increase on decreasing U$_{dd}$. Simultaneously, the intensity of the locally screened feature gets reduced compared to the non-locally screened feature. The ionic character satellites do not show much change for the entire range of U$_{dd}$ investigated here. This result indicates that U$_{dd}$ $\geq$ 4 eV is consistent with the data.

Fig. S8 shows the variation of the calculated Cu 2p X-ray photoemission spectra as a function of the on-site U$_{pp}$, varied from 1 to 6.5 eV. In this case, the calculations start showing changes compared to the experimental spectra for U$_{pp}$ $\geq$ 3 eV. In particular, the energy separation between the main peaks consisting of the dominantly local ( BE = 934.5 eV) and non-local (BE = 933.4 eV) screened states smoothly increase on increasing U$_{pp}$. Simultaneously, the intensity of the non-locally screened feature gets reduced compared to the locally screened feature. The ionic character satellites again show hardly any change for the range of U$_{pp}$ values investigated.
 This result indicates that U$_{pp}$ $<$ 3 eV is consistent with the data.


\end{document}